\documentclass[10pt,conference]{IEEEtran}

\usepackage{cite}
\usepackage{amsmath,amssymb,amsfonts}
\usepackage{array}
\usepackage{booktabs}
\usepackage{colortbl}
\usepackage{graphicx}
\usepackage{multirow}
\usepackage{threeparttable}
\usepackage{textcomp}
\usepackage{tikz}
\usepackage{microtype}
\usepackage[hidelinks]{hyperref}
\usepackage{url}
\usepackage{xcolor}
\usetikzlibrary{calc}

\def\BibTeX{{\rm B\kern-.05em{\sc i\kern-.025em b}\kern-.08em
    T\kern-.1667em\lower.7ex\hbox{E}\kern-.125emX}}

\makeatletter
\def\bstctlcite#1{\@bsphack
  \@for\@citeb:=#1\do{%
    \edef\@citeb{\expandafter\@firstofone\@citeb}%
    \if@filesw\immediate\write\@auxout{\string\citation{\@citeb}}\fi}%
  \@esphack}
\makeatother

\definecolor{barblue}{RGB}{14,103,178}
\definecolor{gridgray}{RGB}{215,220,225}
\definecolor{bordergray}{RGB}{80,80,80}
\definecolor{circlegray}{RGB}{115,115,115}
\definecolor{rqtwfilteredblue}{RGB}{53,89,132}
\definecolor{rqtwexecutedblue}{RGB}{156,186,220}
\definecolor{rqtwlightgray}{RGB}{244,244,244}
\definecolor{rqtwbordergray}{RGB}{195,195,195}
\definecolor{rqtwaxisgray}{RGB}{55,55,55}
\definecolor{rqtwNoReduction}{HTML}{D9DDE3}
\definecolor{rqtwBlueOne}{HTML}{C9DDF2}
\definecolor{rqtwBlueOneDark}{HTML}{BBD4EC}
\definecolor{rqtwBlueTwo}{HTML}{74A9D8}
\definecolor{rqtwBlueThree}{HTML}{3F7FC4}
\definecolor{rqtwBlueFour}{HTML}{1F5A9D}
\definecolor{rqtwLineColor}{HTML}{4A4A4A}
\definecolor{rqtwTextColor}{HTML}{1F1F1F}
\definecolor{BaselineGray}{HTML}{F1F3F5}

\newcommand{\methodllm}[2]{%
    \mbox{%
        \textbf{#1}%
        \kern-0.08em%
        \raisebox{-0.40ex}{\tiny\textbf{#2}}%
    }%
}

\begin{document}

\bstctlcite{IEEEexample:BSTcontrol}

\title{Multi-Perspective Agentic Program Repair via Code Property Graphs and Temporal Execution Graphs}

\author{
\IEEEauthorblockN{
Zhili Huang,
Ling Xu\textsuperscript{*},
and Hongyu Zhang
}
\IEEEauthorblockA{
Chongqing University, Chongqing, China\\
\{huangzhili@stu.cqu.edu.cn, xuling@cqu.edu.cn, hyzhang@cqu.edu.cn\}
}
\thanks{\textsuperscript{*}Corresponding author: Ling Xu (xuling@cqu.edu.cn).}
}

\maketitle

\begin{abstract}
Large language models (LLMs) have improved automated program repair (APR), but two limitations remain. First, raw execution traces are often too large and repetitive to serve as effective model context. Second, repeated patch sampling may produce different implementations without yielding distinct root-cause hypotheses or repair strategies. We present CT-Repair, an agentic APR framework representing static and dynamic evidence as queryable Code Property Graph (CPG) and Temporal Execution Graph (TEG). CT-Repair applies a three-stage filtering pipeline to construct compact TEGs. Three finite-state-machine-guided agents analyze each bug from static, dynamic, and hybrid perspectives and independently produce evidence-grounded repair strategies. A strategy-guided generation procedure instantiates these strategies as candidate patches and uses validation feedback to refine the most promising strategy.

We evaluate CT-Repair on 854 Java bugs from Defects4J v3.0. In the mixed-model configuration, CT-Repair correctly repairs 489 bugs. Under a controlled GPT-5.4-mini configuration, it repairs 388 bugs, 19 and 30 more than ReinFix and RepairAgent, respectively. The union of the three evidence perspectives repairs 99 more bugs than the strongest individual perspective. The filtering pipeline also compacts runtime evidence, with execution filtering narrowing the candidate method scope by 94.85\% on average and behavior filtering further reducing retained runtime records by 55.97\%. These results show that structured runtime evidence and multi-perspective reasoning can improve repair effectiveness without relying solely on a larger patch-generation budget.

\end{abstract}


\section{Introduction}

Automated program repair (APR) aims to automatically generate patches for software bugs~\cite{gazzola2018automatic,huang2024evolving,le2019automated,monperrus2018automatic,zhang2023survey}. Traditional APR techniques, including search-based~\cite{le2011genprog}, constraint-based~\cite{nguyen2013semfix}, and template-based~\cite{liu2019tbar} approaches, have been effective for recurring bug patterns. However, their reliance on predefined search spaces, constraints, or repair patterns limits their ability to adapt to diverse bugs and complex repair scenarios~\cite{xia2024automated}. Learning-based approaches, particularly neural machine translation (NMT)-based APR approaches~\cite{chen2019sequencer,jiang2021cure,li2020dlfix,li2022dear,lutellier2020coconut,meng2022improving,meng2023template,tufano2019empirical,ye2022neural,zhu2021syntax,zhu2023tare}, instead formulate program repair as a translation task. These methods learn from bug-fix pairs (BFPs)~\cite{tufano2019empirical} to translate buggy code into repaired code.

Large language models (LLMs) have recently become an important foundation for APR because of their code understanding and generation capabilities. Existing LLM-based APR methods use multi-turn interaction, test feedback, repair-ingredient retrieval, and tool invocation to support fault analysis and iterative patch refinement~\cite{xia2024automated,zhang2025repair}. More recent agent-based methods further enable LLMs to explore repositories, retrieve relevant context, execute tests, and validate patches within a unified repair workflow~\cite{bouzenia2025repairagent}.

Despite these advances, two limitations remain. The first concerns the use of dynamic execution information. Runtime evidence, including variable states, branch outcomes, method calls, return values, and exception contexts, provides direct observations of program behavior under failing tests and can support the diagnosis of complex logical bugs~\cite{wu2026debugrepair,haque2025towards}. However, raw execution traces are often large, fine-grained, and highly repetitive. Existing methods commonly convert execution logs into text and provide them directly to LLMs~\cite{huang2025dynafix,zhong2024debug}, causing context bloat and possibly burying critical failure evidence in redundant information. When traces exceed the model's context window, retaining only partial execution fragments may lose the call paths and state-propagation process required to explain the failure. The second limitation concerns how heterogeneous evidence is used during repair reasoning. Existing methods may incorporate static program structures, runtime information, error messages, and test feedback~\cite{zhang2025repair,
bouzenia2025repairagent,huang2025dynafix,wu2026runtime}. However, these sources are often combined into a shared context and processed through a single reasoning path to derive a root-cause hypothesis. Candidate patches are then generated and ranked through repeated sampling~\cite{xia2023automated,xia2022less}. Although this process can produce syntactically different patches, the candidates may still rely on the same suspicious locations, root-cause hypotheses, and repair direction. Therefore, patch-level diversity does not necessarily lead to diverse fault analyses or repair strategies. A repair framework should instead exploit the complementary information provided by different evidence sources and explore distinct root-cause hypotheses before candidate-patch generation.

To address these limitations, we present CT-Repair, an agentic APR framework that combines queryable static and dynamic program representations with multi-perspective reasoning. Given a defective program and its failing tests, CT-Repair organizes program evidence at both static and dynamic levels. For static analysis, it uses Joern~\cite{yamaguchi2014modeling} to construct a Code Property Graph (CPG) that captures program structure, control flow, data dependencies, and call relations. For dynamic analysis, CT-Repair instruments the program to collect execution events triggered by failing tests.
It then uses a three-stage filtering pipeline to remove redundant information and constructs a Temporal Execution Graph (TEG) from the remaining runtime evidence.
Rather than directly feeding long execution logs into the model, CT-Repair exposes the CPG and TEG through query interfaces, allowing agents to retrieve evidence relevant to their current fault hypotheses on demand.

Based on these queryable representations, CT-Repair shifts repair diversity from candidate-patch generation to the reasoning stage. The framework runs three agents in parallel, each using a different evidence perspective. The static-oriented agent analyzes program structure and dependencies, while the dynamic-oriented agent examines execution paths and runtime states. The hybrid agent combines static and dynamic evidence to explain how a defect propagates through the program. Guided by a finite-state machine (FSM), the agents independently perform fault understanding, evidence collection, and repair-strategy generation. They derive distinct root-cause hypotheses and corresponding repair strategies before patch generation. CT-Repair then uses a round-robin mechanism to instantiate these strategies as candidate patches and evaluates their quality using compilation and test results. If no plausible patch is produced in the current round, the framework selects the most promising strategy and refines it using validation feedback in the next iteration.

We evaluate CT-Repair on 854 real-world Java bugs from Defects4J v3.0. In the mixed-model configuration, CT-Repair correctly repairs 489 bugs. Under a controlled setting where CT-Repair and the compared methods use GPT-5.4-mini as the base model, CT-Repair correctly repairs 388 bugs, 19 more than the strongest baseline, ReinFix, corresponding to a relative improvement of 5.15\%. These results indicate that CT-Repair improves repair effectiveness under the controlled single-model setting while achieving its highest repair coverage in the mixed-model configuration.

In summary, this paper makes the following contributions:
\begin{itemize}

    \item We propose CT-Repair, an agentic APR framework that introduces reasoning-level diversity through static, dynamic, and hybrid evidence perspectives. Three perspective-specific agents independently derive evidence-grounded root-cause hypotheses and repair strategies before patch generation. A round-robin generation mechanism then instantiates these strategies as candidate patches.
    
    \item We design a Temporal Execution Graph (TEG) to organize dynamic execution evidence. A three-stage filtering process removes unexecuted methods, structurally simple methods, and records disconnected from valid execution flows. Query interfaces allow agents to retrieve evidence relevant to their current fault hypotheses on demand.
    
    \item We evaluate CT-Repair on 854 real-world Java bugs from Defects4J v3.0. The evaluation examines repair effectiveness, runtime-evidence compression, multi-perspective complementarity, and component contributions under controlled model and patch-generation budgets. We also encourage future researchers to leverage our approach to develop more powerful APR tools. Our source code and data are available at: \url{https://anonymous.4open.science/r/CT-Repair-6D29/}.

\end{itemize}

\section{Motivation}

This section examines the limitations of existing LLM-based APR methods in evidence organization and repair reasoning. First, we examine whether raw dynamic execution traces can be used directly as model context. Second, we analyze why repeated patch sampling may introduce implementation-level variation without producing sufficiently diverse root-cause hypotheses. 

\begin{figure*}[t]
    \centering
    \includegraphics[width=\textwidth]{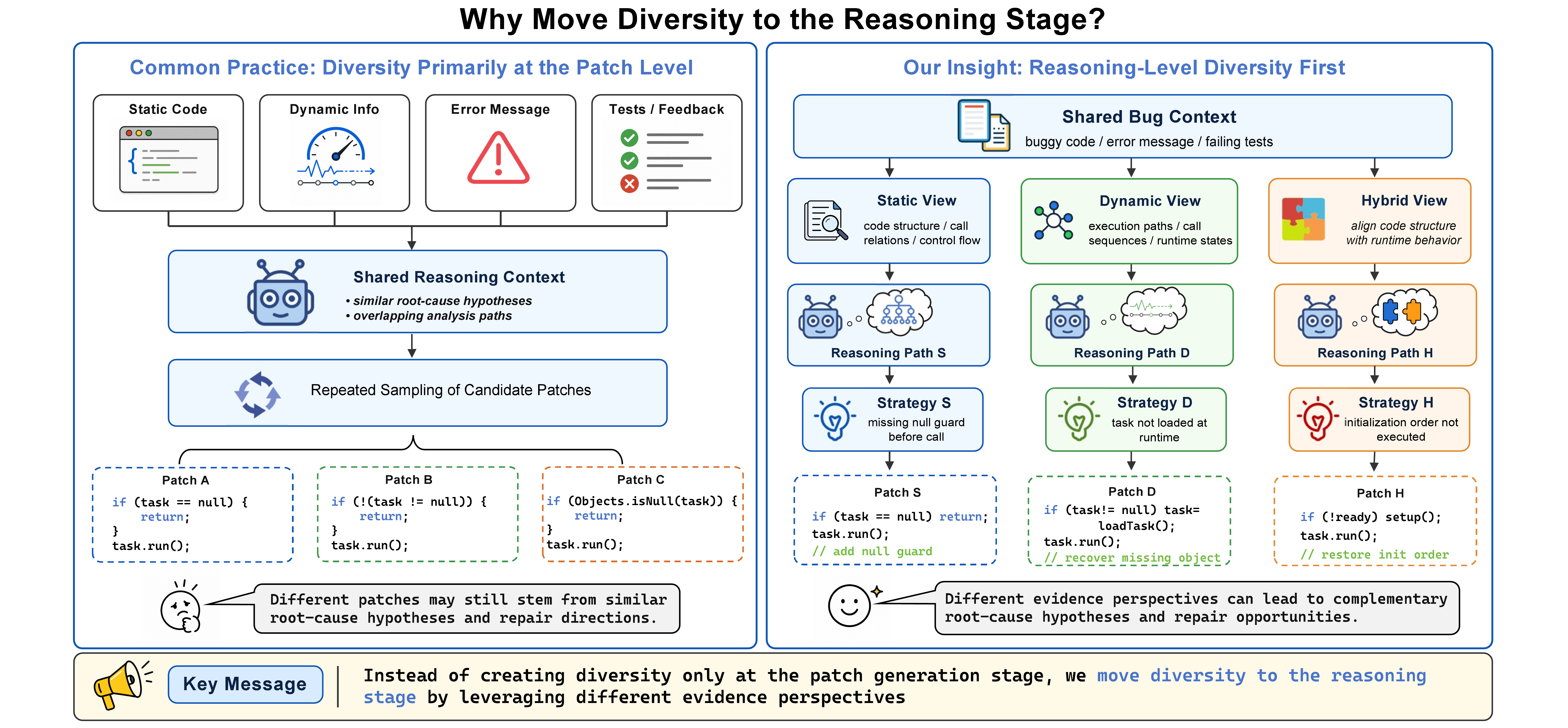}
    \caption{Comparison between patch-level sampling in existing APR workflows and reasoning-level diversity in CT-Repair.}
    \label{fig:motivation}
    
\end{figure*}

\subsection{Dynamic Execution Evidence: Informative but Difficult to Use}

Recent LLM-based APR methods use dynamic execution information to characterize runtime behavior triggered by failing tests. However, execution traces are commonly serialized as linear text and directly included in the model context, rather than organized as structured representations of program behavior. To assess the practicality of using raw traces in this manner, we conduct a motivational study on 17 Defects4J projects. 
Because collecting and processing complete execution traces is expensive, we proportionally sample 100 bugs according to project-level bug distributions. As shown in Table~\ref{tab:motivation_trace_redundancy}, each bug produces 2.38 million runtime events on average. Of these events, only 13.15K occur in methods modified by the corresponding developer patches, accounting for 0.55\% on average. In addition, 99.95\% of the runtime events are repetitive under our event-signature definition.

These results reveal a core tension in representing dynamic execution evidence. Complete traces preserve interprocedural call paths, branch outcomes, and state propagation, but their size and redundancy make them difficult to use directly as model context. By contrast, retaining only a localized subset of events yields a more compact representation but may discard the key execution context needed for diagnosis. Therefore, dynamic execution evidence requires a structured, compact, and queryable organization that reduces redundant records while preserving the temporal relations and interprocedural context needed for diagnosis, and supports repair agents in retrieving hypothesis-relevant evidence on demand.

\begin{table}[t]
    \centering
    \caption{Project-level statistics of raw dynamic execution traces.}
    \label{tab:motivation_trace_redundancy}
    \setlength{\tabcolsep}{2.4pt}
    \renewcommand{\arraystretch}{1.08}
    \begin{threeparttable}
    \resizebox{\columnwidth}{!}{%
    \begin{tabular}{@{}lccccc@{}}
        \toprule
        \multirow{2}{*}{\textbf{Project}} &
        \multirow{2}{*}{\textbf{\#Bugs}} &
        \textbf{Full} &
        \textbf{Developer-Patch} &
        \textbf{Event} &
        \textbf{Repetitive} \\
        & & \textbf{Events} & \textbf{Method Events} & \textbf{Ratio} & \textbf{Event Ratio} \\
        \midrule
        Chart           & 4  & 666.40K & 867     & 0.13\%  & 98.72\% \\
        Cli             & 5  & 16.85K  & 431     & 2.56\%  & 90.56\% \\
        Closure         & 14 & 206.50M & 10.71K  & 0.01\%  & 99.98\% \\
        Codec           & 4  & 17.12K  & 11.70K  & 68.34\% & 95.81\% \\
        Collections     & 5  & 19.09K  & 99      & 0.52\%  & 97.75\% \\
        Compress        & 6  & 59.30K  & 1.65K   & 2.78\%  & 92.72\% \\
        Csv             & 4  & 8.14K   & 150     & 1.84\%  & 89.47\% \\
        Gson            & 4  & 6.03K   & 1.16K   & 19.15\% & 89.24\% \\
        JacksonCore     & 4  & 4.06M   & 1.09M   & 26.82\% & 99.92\% \\
        JacksonDatabind & 10 & 1.08M   & 862     & 0.08\%  & 97.50\% \\
        JacksonXml      & 3  & 4.43K   & 384     & 8.68\%  & 77.27\% \\
        Jsoup           & 7  & 19.17M  & 158.33K & 0.83\%  & 99.99\% \\
        JxPath          & 4  & 1.66M   & 11.72K  & 0.70\%  & 99.35\% \\
        Lang            & 7  & 517.81K & 15.82K  & 3.06\%  & 99.80\% \\
        Math            & 10 & 4.26M   & 10.65K  & 0.25\%  & 99.93\% \\
        Mockito         & 5  & 26.57K  & 350     & 1.32\%  & 96.46\% \\
        Time            & 4  & 52.79K  & 996     & 1.89\%  & 95.62\% \\
        \midrule
        \textbf{Bug Avg.} &
        \textbf{100} &
        \textbf{2.38M} &
        \textbf{13.15K} &
        \textbf{0.55\%} &
        \textbf{99.95\%} \\
        \bottomrule
    \end{tabular}%
    }

    \vspace{2pt}

    \begin{minipage}{\columnwidth}
    \scriptsize
    \textit{Note:}
    Event Ratio denotes the share of events in developer-modified functions. Repetitive Event Ratio is \(1 - N_u / N\), where \(N_u\) and \(N\) denote unique event signatures and total runtime events.
    \end{minipage}

    \end{threeparttable}

\end{table}

\subsection{From Patch-Level Diversity to Reasoning-Level Diversity}

Repeated candidate-patch sampling does not necessarily lead to diverse fault diagnoses. As illustrated on the left side of Fig.~\ref{fig:motivation}, a common LLM-based APR workflow combines source code, dynamic execution evidence, error messages, and test feedback into a shared context, and then samples multiple candidate patches~\cite{xia2023automated,xia2022less}. This process may produce syntactically different patches while preserving the same underlying root-cause hypothesis. Consider a null-pointer failure. Different samples may introduce an explicit null check, rewrite the corresponding condition, or invoke a utility method that performs an equivalent check. Although these patches differ at the code level, they share the same hypothesis that a null check is missing before the call. Repeated sampling therefore explores alternative implementations of the same repair direction, rather than alternative explanations of the failure.

As shown on the right side of Fig.~\ref{fig:motivation}, our key idea is to build relatively independent reasoning paths from static, dynamic, and hybrid perspectives instead of merging all information into one process. The static-oriented agent focuses on code structure and dependencies, the dynamic-oriented agent on actual execution paths and runtime states, and the hybrid-oriented agent on aligning structure with behavior. Each agent independently formulates and validates a root-cause hypothesis before producing a repair strategy. As a result, candidate patches can be generated from different fault explanations and modification directions, rather than only from stochastic variations of a shared reasoning process. CT-Repair thus introduces diversity at the reasoning stage while retaining implementation-level diversity through subsequent strategy-guided patch generation.

\section{Approach}

\subsection{Overview}
\begin{figure*}[t]
    \centering
    \includegraphics[width=\textwidth]{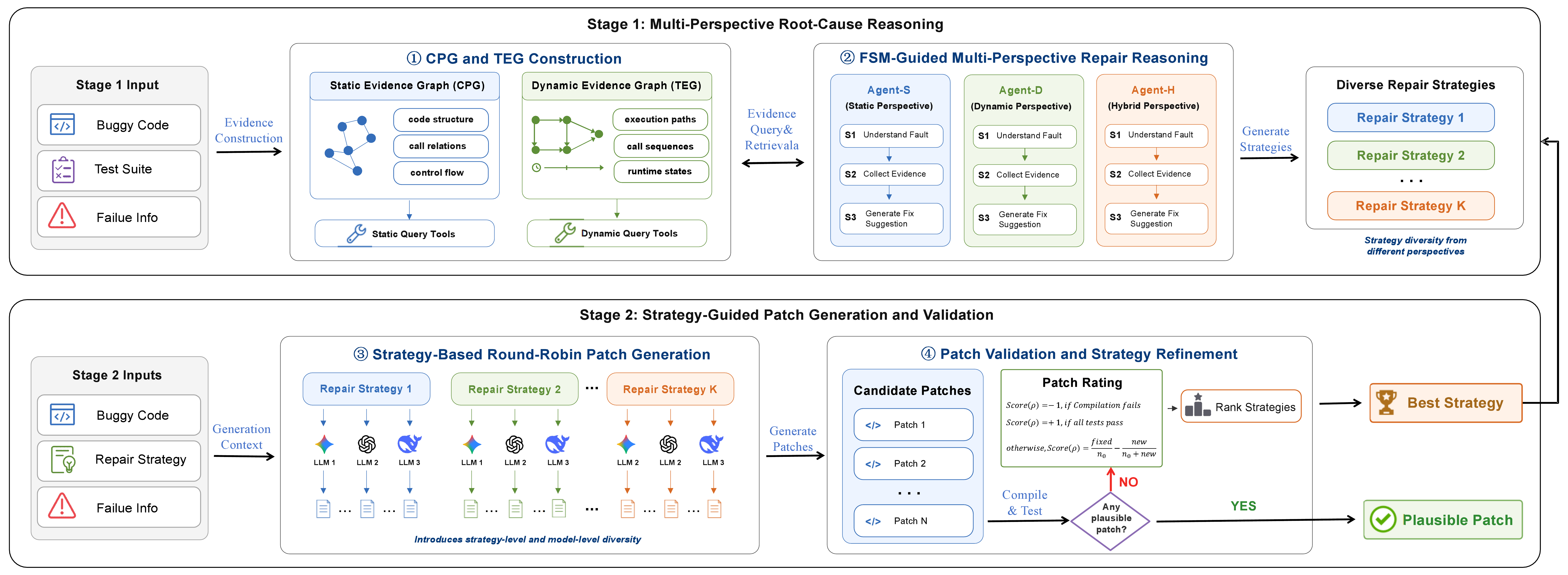}
    \caption{Overall workflow of CT-Repair.}
    \label{fig:overall_framework}
\end{figure*}

This section introduces the proposed CT-Repair framework. The framework generates candidate repairs through four components organized into two stages, as shown in Fig.~\ref{fig:overall_framework}. The first stage constructs queryable static and dynamic evidence and uses multi-perspective agents to derive repair strategies. The second stage instantiates these strategies as candidate patches, validates the patches, and uses validation feedback to refine promising strategies.

In the first stage, CT-Repair constructs a CPG for the target codebase and a TEG from the executions triggered by failing tests. Query interfaces over the two graphs provide static and dynamic evidence on demand. Three finite-state-machine-guided agents then analyze the same bug from static, dynamic, and hybrid perspectives. Each agent independently formulates and validates a root-cause hypothesis before deriving a corresponding repair strategy.

In the second stage, CT-Repair uses strategy-based round-robin scheduling to instantiate each repair strategy with multiple LLMs. The resulting candidate patches are compiled and tested. If no plausible patch is found, CT-Repair ranks the repair strategies based on the validation outcomes of their candidate patches. It then returns feedback from the best-performing patch of the highest-scoring strategy to the originating agent for further refinement.

\subsection{CPG and TEG Construction}
 
This stage organizes static program structure and runtime behavior into two complementary graph representations. Supplying an LLM with complete repository code or raw execution logs can introduce substantial irrelevant and repetitive information, making relevant evidence harder to identify in long contexts~\cite{liu2024lost}. CT-Repair therefore constructs a CPG and a TEG and exposes query interfaces over both graphs. These interfaces allow the agents to retrieve evidence relevant to their current hypotheses without processing the complete program or execution trace in every model call. Table~\ref{tab:analysis_tools} summarizes the static and dynamic analysis tools available to the agents.

\begin{table}[t]
    \centering
    \caption{Static and dynamic analysis tools available to CT-Repair agents.}
    \label{tab:analysis_tools}
    \resizebox{\columnwidth}{!}{%
    \begin{tabular}{@{}clll@{}}
        \toprule
        \textbf{Source} & \textbf{Tool Name} & \textbf{Type} & \textbf{Description} \\
        \midrule
        \multirow{8}{*}{CPG}
        & static\_identify\_variable & Line-Level & Finds variable definitions, references, and types in a file \\
        & static\_trace\_method\_usage & Method-Level & Retrieves static call sites of a target method \\
        & static\_analyze\_method\_details & Method-Level & Extracts parameters, returns, calls, assignments, and locals \\
        & static\_analyze\_method\_control\_flow & Method-Level & Extracts control structures within a target method \\
        & static\_get\_method\_javadoc & Method-Level & Retrieves the Javadoc comment of a target method \\
        & static\_get\_class\_structure & Class-Level & Retrieves fields and declared methods of a target class \\
        & static\_get\_class\_javadoc & Class-Level & Retrieves the Javadoc comment of a target class \\
        & static\_get\_imports & File-Level & Retrieves import declarations from a target file \\
        \midrule
        \multirow{6}{*}{TEG}
        & dynamic\_method\_macro & Method-Level & Summarizes return variants, callers, and callees \\
        & dynamic\_method\_execution\_summary & Method-Level & Groups executions by arguments, returns, and sequences \\
        & dynamic\_sequences\_info & Invocation-Level & Retrieves runtime events for selected executions \\
        & dynamic\_chronological\_flow & Invocation-Level & Traces timestamp-ordered events from one execution \\
        & dynamic\_call\_context & Invocation-Level & Retrieves upstream and downstream runtime call context \\
        & dynamic\_class\_execution\_paths & Class-Level & Summarizes runtime paths among methods in a class \\
        \bottomrule
    \end{tabular}%
    }
\end{table}

\textbf{CPG Construction.} 
The CPG captures static structures and potential dependencies of the target codebase. CT-Repair uses Joern~\cite{yamaguchi2014modeling} to construct a unified representation of abstract syntax, control flow, data flow, and method-call relations.
CT-Repair exposes CPG queries at four levels of granularity. Line-level queries locate variable definitions, references, and types. Method-level queries retrieve parameters, return information, assignments, control structures, and static call sites. Class-level queries provide declared fields, methods, and related Javadoc documentation. File-level queries retrieve import declarations that indicate dependencies on external components. These queries allow agents to examine the structural context of suspicious code and narrow the search space for potential repair locations.

\textbf{TEG Construction.} The TEG captures the runtime behavior observed during failing executions. CT-Repair first narrows the dynamic tracing scope and instruments retained methods at the bytecode level. The instrumentation records runtime events, including method calls, parameters and return values, variable updates, and branch outcomes. CT-Repair then organizes these events into a TEG, where nodes represent method-call instances and edges represent actual calls and temporal order. Each event keeps its timestamp, sequence number, and call index, allowing agents to query specific methods, call instances, and runtime states along the actual execution trace.

\textbf{Dynamic Trace Filtering.} 
Raw execution traces contain many records that provide limited additional evidence for fault diagnosis or are repeatedly generated by structurally simple methods. CT-Repair applies a three-stage filtering pipeline to control the size of the resulting TEG. First, execution filtering (EF) uses JaCoCo coverage to exclude methods that are not executed by any failing test. Second, structural filtering (SF) analyzes the AST and excludes empty methods, trivial getters and setters, and simple delegators. The semantics of these methods can generally be recovered from source code, whereas instrumenting frequently invoked instances of them may generate large numbers of repetitive runtime events. EF and SF therefore jointly determine the scope of bytecode instrumentation. After the failing tests have been executed, behavior filtering (BF) removes isolated records and call relations that cannot be connected to a valid invocation chain in the reconstructed execution. Such records lack sufficient caller-callee or temporal context for analyzing fault propagation. CT-Repair constructs the final TEG from the remaining records, retaining the observed call relations, event order, and runtime states of the failing executions.

\subsection{FSM-Guided Multi-Perspective Repair Reasoning}
\label{sec:fsm_reasoning}
Using the query interfaces over the CPG and TEG, CT-Repair uses an FSM to guide three agents in analyzing bugs from different evidence perspectives and generating repair strategies.

\textbf{Multi-Perspective Root-Cause Reasoning.} CT-Repair runs three agents in parallel. All agents receive the same basic bug context, including the buggy code, triggering tests, and
failure messages, but prioritize different evidence. Agent-S primarily uses CPG-based tools to examine program structure, static call relations, and control- and data-flow dependencies. Agent-D primarily uses TEG-based tools to inspect observed execution paths, call sequences, branch outcomes, and runtime states. Agent-H uses both tool sets to relate runtime anomalies to program structure and trace state propagation across methods. Each agent independently formulates a root-cause hypothesis and derives one initial repair strategy. This design introduces diversity at the reasoning stage rather than relying only on repeated patch sampling from a shared analysis context. CT-Repair supports both single-model and mixed-model configurations. In the mixed-model configuration, different LLMs are assigned to the three agents, and the model-to-perspective assignments are rotated across bugs to avoid consistently associating one perspective with a particular model~\cite{snell2025scaling}.

\begin{figure*}[t]
    \centering
    \includegraphics[width=0.95\textwidth]{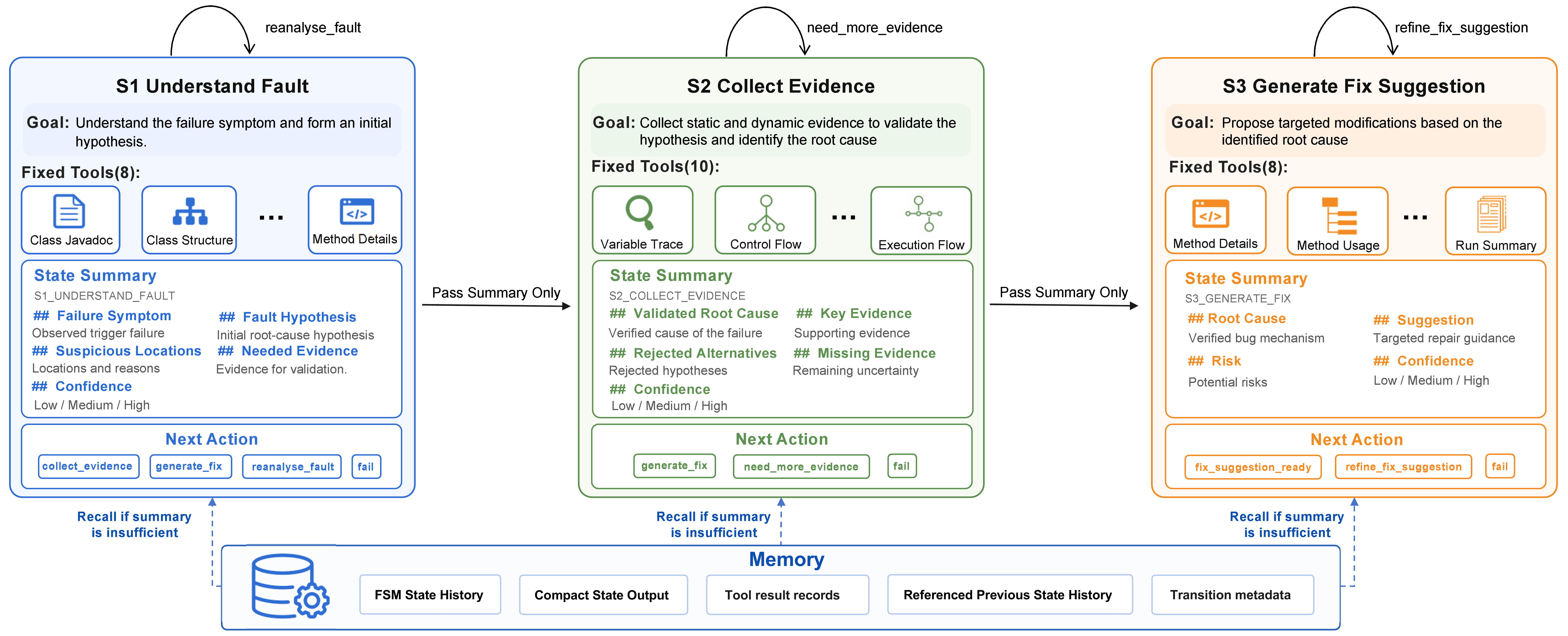}
    \caption{FSM-guided multi-perspective repair reasoning.}
    \label{fig:fsm_reasoning}
\end{figure*}

\textbf{FSM-Guided Reasoning.} CT-Repair uses an FSM to structure the analysis process and regulate tool use. As shown in Fig.~\ref{fig:fsm_reasoning}, the FSM contains three analysis states. Each state defines an analysis objective, a structured output format, and valid transitions. The tools available to an agent are determined jointly by its evidence perspective and current state. An agent advances after producing the required state output; otherwise, it follows a self-loop and continues the current analysis.

\begin{itemize}
    \item \emph{\texttt{understand\_fault}.} The agent examines triggering tests, failure messages, and buggy code to characterize the failure, identify suspicious locations, and
    formulate an initial root-cause hypothesis. If the analysis remains incomplete, the agent follows \texttt{reanalyze\_fault} and stays in the current state. Otherwise, it transitions to \texttt{collect\_evidence}.

    \item \emph{\texttt{collect\_evidence}.}  The agent invokes the tools available to its evidence perspective and uses the retrieved evidence to support, revise, or reject the current hypothesis. If further evidence is needed, it follows \texttt{need\_more\_evidence} and continues the analysis. Once the evidence sufficiently supports a root-cause explanation, the agent transitions to \texttt{generate\_fix\_suggestion}.

    \item \emph{\texttt{generate\_fix\_suggestion}.} The agent derives a structured repair strategy from the supported root-cause hypothesis and collected evidence. The strategy identifies
    the target methods, describes the intended changes, and records
    potential risks or uncertainties. If the strategy is incomplete, the agent follows \texttt{refine\_fix\_suggestion} and revises it. Once the strategy satisfies the output schema, the agent emits \texttt{fix\_suggestion\_ready}, transitions to \texttt{done}, and forwards the strategy to patch generation.
\end{itemize}

\textbf{Summary-Based State Management.} CT-Repair uses summary-based state transfer to control context growth during multi-step reasoning. At each state transition, it ends the current LLM interaction and passes a structured state summary to the next state. The summary records the state-specific findings, unresolved issues, and selected transition action. The interaction history and tool outputs associated with each state are stored separately. If the summary lacks sufficient detail, the agent can retrieve relevant records from earlier states on demand. This design avoids repeatedly including the full reasoning history in subsequent model calls while preserving access to previously collected evidence.

\subsection{Strategy-Based Round-Robin Patch Generation}
After the three agents produce their repair strategies, CT-Repair instantiates each strategy as concrete candidate patches. A strategy is represented as \(\langle\textit{Root Cause}, \textit{Suggestion}\rangle\), where \textit{Root Cause} provides an evidence-supported explanation of the failure and \textit{Suggestion} specifies the target locations and intended changes. CT-Repair supplies the strategy together with the required bug context to the patch-generation models. Candidate generation is therefore conditioned on an explicit diagnosis and
modification plan rather than on the raw bug context alone.

As shown in Fig.~\ref{fig:overall_framework}, CT-Repair uses strategy-guided round-robin scheduling to cycle through multiple patch-generation models for each strategy. If a strategy is processed by \(r\) models and each model generates at most \(k\) candidates, the strategy produces at most \(r \times k\) patches. This procedure introduces implementation-level variation while keeping the underlying root-cause hypothesis and repair direction fixed.

Each candidate patch retains an identifier for the strategy from which it was generated. CT-Repair can therefore associate compilation and test outcomes with the corresponding root-cause hypothesis and modification plan. These associations are subsequently used to evaluate the strategies and select one for further refinement.

\subsection{Patch Validation and Strategy Refinement}

\textbf{Patch Validation.} CT-Repair validates the candidate patches generated in the current iteration in batches. Each patch is applied to the buggy program and evaluated through compilation and testing. A patch that compiles and passes all available tests is marked as \texttt{PLAUSIBLE}. Only when no \texttt{PLAUSIBLE} patch exists, CT-Repair records the validation outcome of each candidate and proceeds to patch scoring and strategy selection.

\textbf{Patch Scoring and Strategy Refinement.} Let \(F_0\) be the set of tests that fail on the original buggy program, and let \(F_p\) denote the set of tests that fail after applying patch \(p\). The set \(F_0 - F_p\) represents the original failing tests failures eliminated by \(p\), whereas \(F_p - F_0\) represents new failures introduced by the patch. The score of patch \(p\) is defined as follows:

\begin{equation}
\resizebox{0.88\columnwidth}{!}{$\displaystyle
\operatorname{Score}(p)=
\begin{cases}
-1, & \text{if } p \text{ fails to compile},\\
+1, & \text{if } p \text{ passes all tests},\\
\frac{|F_0 - F_p|}{|F_0|}
-
\frac{|F_p - F_0|}{|F_0|+|F_p - F_0|},
& \text{otherwise}.
\end{cases}
$}
\end{equation}

For a compilable but non-plausible patch, the first term rewards the elimination of original test failures, while the second term penalizes newly introduced failures. The score therefore provides more fine-grained feedback than a binary compilation or test outcome. Let \(P_s\) denote the candidate patches generated from strategy \(s\). CT-Repair evaluates the strategy using the mean score of its candidate patches:

\begin{equation}
Q(s)=\frac{1}{|P_s|}\sum_{p\in P_s}\operatorname{Score}(p).
\end{equation}

If \(Q(s) \leq 0\) for every strategy, CT-Repair terminates the repair process because the current validation results provide no positive signal for further refinement. Otherwise, it selects the strategy with the highest \(Q(s)\) and identifies the highest-scoring patch generated from that strategy. CT-Repair extracts feedback from this patch, including its compilation status, remaining failing tests, newly introduced failures, and code changes. This feedback is returned to the agent that generated the selected strategy, enabling it to re-examine the root cause and refine the repair suggestion. Subsequent iterations focus only on this agent's strategy, concentrating the limited budget on the most promising repair direction.

\section{Experiment Setup}

\subsection{Research Questions}

Our study addresses the following research questions (RQs):

\begin{itemize}
    \item \textbf{RQ1:} How does CT-Repair compare with state-of-the-art APR methods?

    \item \textbf{RQ2:} How effectively does TEG reduce execution data?

    \item \textbf{RQ3:} Do multi-perspective agents improve reasoning diversity and repair complementarity?

    \item \textbf{RQ4:} How does each major component contribute to CT-Repair's effectiveness?
\end{itemize}

\subsection{Benchmarks}

\textbf{Defects4J.}
We conduct the primary evaluation on the widely used Defects4J benchmark~\cite{just2014defects4j}. We use Defects4J v3.0\footnote{\url{https://github.com/rjust/defects4j/tree/v3.0.0}}, which contains 854 real bugs from 17 Java projects. Following prior studies~\cite{zhang2025repair,yin2024thinkrepair}, we classify bugs according to the number of methods modified by the developer patch. The benchmark contains 483 single-function bugs and 371 multi-function bugs. Defects4J provides buggy and fixed versions together with test suites, enabling reproducible evaluation and fair comparison with existing APR techniques.

\subsection{Compared Techniques}

To evaluate repair performance, we compare CT-Repair with two recent and representative LLM-based agentic APR baselines.

\textbf{RepairAgent}~\cite{bouzenia2025repairagent}:
proposes an autonomous LLM-based APR agent that combines dynamic prompting, an FSM, and repair tools to collect bug information, search for repair ingredients, generate patches, and validate them.

\textbf{ReinFix}~\cite{zhang2025repair}:
presents an LLM-based APR method based on repair-ingredient search. During reasoning, it retrieves internal repair ingredients such as variable definitions to support root-cause analysis. During patch generation, it retrieves external repair ingredients from historical bug fixes to improve patch-generation accuracy.

For fair comparison, we reproduce both baselines and rerun them under a unified experimental setting. Specifically, we replace their original large language models with the same model backbones used by CT-Repair to reduce the influence of model-capability differences. All methods are evaluated on the same Defects4J bug set and use the same patch-validation workflow and decision criteria for plausible and correct patches.

\subsection{Evaluation Metrics}
We adopt the standard evaluation metrics used in prior APR studies. A plausible patch is defined as a patch that passes all provided unit tests, while a correct patch is a plausible patch that is semantically equivalent to the developer's fix. To determine correctness, two authors independently inspected all plausible patches and resolved disagreements through adjudication until consensus was reached, ensuring that the correctness metric reflects semantic equivalence rather than mere test passing.

\subsection{Implementation}

CT-Repair is implemented with LangGraph and uses three language models: GPT-5.4-mini, Gemini-3-flash, and DeepSeek-v4-flash. These models are used for both multi-perspective agent reasoning and patch generation. We set the temperature of reasoning agents to 0 for stable reasoning and the patch-generation temperature to 1 to increase candidate-patch diversity. We also disable the model reasoning budget to control additional reasoning overhead and ensure consistency across model settings. In fault localization (FL), to avoid additional biases introduced by the FL tool, we follow recent works~\cite{bouzenia2025repairagent,zhang2025repair} under conditions of perfect fault localization.

CT-Repair uses three parallel repair agents, each generating at most one repair strategy per bug from a different perspective. In the first iteration, each strategy is instantiated by three LLMs, yielding at most 3 $\times$ 3 candidate patches. If no plausible patch is found, CT-Repair refines only the best-performing strategy and instantiates it with the same three LLMs, generating three additional patches. Thus, the maximum patch budget for one bug is 3 $\times$ 3 + 3 candidate patches.

To prevent ineffective loops, we limit the number of LangGraph node transitions in a single run to 50. In each FSM analysis state, an agent can call analysis tools at most four times. These limits control tool-call cost and runtime while improving experimental reproducibility.

\section{Evaluation}

\subsection{RQ1: Repair Effectiveness}
\label{sec:rq1}

To address RQ1, we evaluate CT-Repair under different base-model configurations. In single-model settings, all three agents use the same LLM. In the mixed-model configuration, the three agents use different LLMs, with model assignments rotated across bugs. This design avoids binding specific informational perspectives to any single model. We then compare CT-Repair with both ReinFix and RepairAgent under the same GPT-5.4-mini setting, considering both repair effectiveness and sampling budget.

\textbf{Repair Effectiveness and Complementarity across Models.} As shown in Table~\ref{tab:llm_settings}, CT-Repair achieves strong repair effectiveness under all three single-model settings. When using Gemini-3-flash, DeepSeek-v4-flash, and GPT-5.4-mini as the base LLM, CT-Repair correctly repairs 460, 410, and 388 bugs. The mixed-model configuration achieves the best overall result, generating 573 plausible patches, including 489 correct ones. Compared with the best-performing single-model setting using Gemini-3-flash, the mixed-model configuration repairs 29 additional bugs (489 vs. 460), corresponding to a 6.30\% improvement, and achieves gains of 19.27\% and 26.03\% over DeepSeek-v4-flash and GPT-5.4-mini, respectively. These results indicate that different LLMs provide complementary bug understanding and patch generation capabilities, enabling CT-Repair to repair bugs missed by individual single-model configurations.

\textbf{Effectiveness and Efficiency Trade-off.} Different model configurations show different trade-offs between repair effectiveness and computational cost. Gemini-3-flash achieves the shortest average runtime of 112.78 s, while GPT-5.4-mini has the lowest average token consumption of 134.78K. The mixed-model configuration achieves the best repair effectiveness with an average cost of 178.35K tokens and 158.90 s per bug, both within the ranges observed across the three single-model settings.

\begin{table}[t]
    \centering
    \caption{Repair results (correct fixes / plausible fixes) and cost across LLM settings.}
    \label{tab:llm_settings}

    \renewcommand{\arraystretch}{1.2}
    \setlength{\tabcolsep}{3pt}

    \resizebox{\columnwidth}{!}{%
    \begin{tabular}{@{}lccccc@{}}
        \toprule
        \textbf{LLM}
        & \textbf{Multi-function}
        & \textbf{Single-function}
        & \textbf{All}
        & \textbf{Avg. Tokens (K)}
        & \textbf{Avg. Time (s)} \\
        \midrule
        Gemini-3-flash
        & 124/156
        & 336/395
        & 460/551
        & 153.99
        & \textbf{112.78} \\

        DeepSeek-v4-flash
        & 103/141
        & 307/364
        & 410/505
        & 212.83
        & 204.94 \\

        GPT-5.4-mini
        & 93/115
        & 295/346
        & 388/461
        & \textbf{134.78}
        & 305.19 \\

        Mixed-model
        & \textbf{130/162}
        & \textbf{359/411}
        & \textbf{489/573}
        & 178.35
        & 158.90 \\
        \bottomrule
    \end{tabular}
    }

\end{table}

\textbf{Comparison with Representative LLM-based agentic APR baselines.} As shown in Table~\ref{tab:repair_performance_comparison}, under the same GPT-5.4-mini setting, CT-Repair correctly repairs 388 bugs and generates 461 plausible patches, outperforming both ReinFix and RepairAgent. Compared with ReinFix, CT-Repair repairs 19 more bugs (388 vs. 369, +5.15\%). Compared with RepairAgent, CT-Repair repairs 30 more bugs (388 vs. 358, +8.38\%). In addition, CT-Repair uses a sampling budget of only \(3 \times 3 + 3\), which is smaller than ReinFix's \(3 \times 3 \times 5\) and RepairAgent's 117 attempts. Overall, CT-Repair not only achieves better repair effectiveness but also outperforms representative existing methods with fewer samples, further validating the effectiveness of its framework design.

\begin{table}[t]
    \centering
    \caption{Repair results (correct fixes / plausible fixes) for CT-Repair and baselines on Defects4J v3.0 under GPT-5.4-mini.}
    \label{tab:repair_performance_comparison}

    \renewcommand{\arraystretch}{1.12}

    \resizebox{\columnwidth}{!}{
    \begin{tabular}{@{}lcccc@{}}
        \toprule
        \textbf{Project}
        & \textbf{\#Bugs}
        & \textbf{CT-Repair}
        & \textbf{ReinFix}
        & \textbf{RepairAgent} \\

        \textnormal{Sampling Times}
        & --
        & $3 \times 3 + 3$
        & $3 \times 3 \times 5$
        & 117 \\
        \midrule

        Chart           & 26  & \textbf{19/20} & 18/19  & 14/17 \\
        Cli             & 39  & \textbf{18/22} & 16/19  & 15/19 \\
        Closure         & 174 & \textbf{49/63} & 46/59  & 40/53 \\
        Codec           & 18  & 10/12  & \textbf{13/15} & 12/14 \\
        Collections     & 28  & 13/14  & 11/15  & \textbf{16/22} \\
        Compress        & 47  & 23/26  & \textbf{25/28} & 18/25 \\
        Csv             & 16  & 12/14  & 10/10  & \textbf{13/13} \\
        Gson            & 18  & 11/11  & 11/12  & \textbf{13/13} \\
        JacksonCore     & 26  & \textbf{12/14} & \textbf{12/14} & 11/12 \\
        JacksonDatabind & 110 & \textbf{48/61} & 45/59  & 43/48 \\
        JacksonXml      & 6   & 1/1    & \textbf{2/2} & \textbf{2/2} \\
        Jsoup           & 93  & \textbf{46/49} & 43/45  & 44/49 \\
        JxPath          & 22  & 6/7    & 2/6    & \textbf{8/11} \\
        Lang            & 61  & \textbf{42/48} & 40/46  & 33/42 \\
        Math            & 106 & \textbf{55/71} & 50/67  & \textbf{55/71} \\
        Mockito         & 38  & 17/18  & \textbf{18/18} & 10/13 \\
        Time            & 26  & 6/10   & 7/9    & \textbf{11/12} \\

        \midrule
        \textbf{Total}
        & 854
        & \textbf{388/461}
        & 369/443
        & 358/436 \\
        \bottomrule
    \end{tabular}
    }

\end{table}

\noindent\textbf{Answer to RQ1.}
CT-Repair achieves the best overall repair effectiveness in the mixed-model configuration and outperforms ReinFix and RepairAgent under the same GPT-5.4-mini setting with a smaller sampling budget.

\subsection{RQ2: Dynamic Information Compression}

To answer RQ2, we evaluate whether CT-Repair's three-stage filtering mechanism reduces the size and redundancy of dynamic execution information. Specifically, we compare the scales of methods, events, records, and traces before and after each filtering stage. Among the 854 bugs, 827 support unfiltered-trace collection; the remaining 27 are excluded from RQ2 because their unfiltered traces are too large for stable collection.

\begin{figure}[t]
    \centering
    \resizebox{\columnwidth}{!}{%
    \begin{tikzpicture}[
        x=1cm,
        y=1cm,
        line cap=round,
        line join=round,
        font=\normalfont
    ]

    \fill[rqtwfilteredblue] (3.45,8.85) rectangle (3.97,9.18);
    \node[
        anchor=west,
        font=\fontsize{10.5}{12}\selectfont
    ] at (4.00,9.015)
    {Removed Methods};

    \fill[rqtwexecutedblue] (7.25,8.85) rectangle (7.77,9.18);
    \node[
        anchor=west,
        font=\fontsize{10.5}{12}\selectfont
    ] at (7.80,9.015)
    {Remaining Methods};

    \def\xstart{2.25}
    \def\xend{13.05}
    \def\barwidth{10.80}
    \def\barheight{1.28}

    \newcommand{\drawfilterbar}[6]{%
        \pgfmathsetmacro{\filteredwidth}{#2/100*\barwidth}
        \pgfmathsetmacro{\executedwidth}{#3/100*\barwidth}
        \pgfmathsetmacro{\splitx}{\xstart+\filteredwidth}
        \pgfmathsetmacro{\bartop}{#1+\barheight}
        \pgfmathsetmacro{\darklabelx}{\xstart+\filteredwidth/2}
        \pgfmathsetmacro{\lightlabelx}{\splitx+\executedwidth/2}

        \fill[rqtwfilteredblue]
            (\xstart,#1) rectangle (\splitx,\bartop);

        \fill[rqtwexecutedblue]
            (\splitx,#1) rectangle (\xend,\bartop);

        \draw[line width=0.65pt, color=rqtwaxisgray]
            (\xstart,#1) rectangle (\xend,\bartop);

        \draw[line width=0.55pt, color=rqtwaxisgray]
            (\splitx,#1) -- (\splitx,\bartop);

        \node[
            anchor=center,
            text=white,
            font=\bfseries\fontsize{12}{14}\selectfont
        ] at (\darklabelx,{#1+\barheight/2})
        {#2\%};

        \draw[line width=0.45pt, color=rqtwaxisgray]
            (\lightlabelx,\bartop) -- (\lightlabelx,{\bartop+0.15});

        \node[
            anchor=south,
            text=black,
            font=\bfseries\fontsize{12}{14}\selectfont
        ] at (\lightlabelx,{\bartop+0.18})
        {#3\%};

        \node[
            anchor=north,
            font=\fontsize{10.2}{12}\selectfont
        ] at ({(\xstart+\xend)/2},{#1-0.19})
        {#4 $\longrightarrow$ #5 methods};

    }

    \node[
        anchor=center,
        align=center,
        font=\fontsize{10}{15}\selectfont
    ] at (0.95,7.72)
    {\textbf{Single-function}\\[-1pt]\fontsize{11}{11.5}\selectfont (466 bugs)};

    \node[
        anchor=center,
        align=center,
        font=\fontsize{10}{15}\selectfont
    ] at (0.95,5.27)
    {\textbf{Multi-function}\\[-1pt]\fontsize{11}{11.5}\selectfont (361 bugs)};

    \node[
        anchor=center,
        align=center,
        font=\fontsize{10}{15}\selectfont
    ] at (0.95,2.82)
    {\textbf{Overall}\\[-1pt]\fontsize{11}{11.5}\selectfont (827 bugs)};

    \drawfilterbar
        {7.08}
        {95.12}
        {4.88}
        {3,995.77}
        {194.92}
        {95.12}

    \drawfilterbar
        {4.63}
        {94.48}
        {5.52}
        {3,846.45}
        {212.24}
        {94.48}
    
    \drawfilterbar
        {2.18}
        {94.85}
        {5.15}
        {3,930.59}
        {202.48}
        {94.85}

    \draw[line width=0.70pt, color=rqtwaxisgray]
        (\xstart,1.35) -- (\xstart,8.72);

    \draw[line width=0.70pt, color=rqtwaxisgray]
        (\xstart,1.35) -- (\xend,1.35);

    \foreach \value in {0,20,40,60,80,100} {
        \pgfmathsetmacro{\tickx}{\xstart+\value/100*\barwidth}

        \draw[line width=0.65pt, color=rqtwaxisgray]
            (\tickx,1.35) -- (\tickx,1.18);

        \node[
            anchor=north,
            font=\fontsize{10.5}{12}\selectfont
        ] at (\tickx,1.03)
        {\value\%};
    }

    \node[
        anchor=center,
        font=\fontsize{11.8}{13}\selectfont
    ] at ({(\xstart+\xend)/2},0.38)
    {Percentage of Methods};

    \end{tikzpicture}%
    }
    \caption{Reduction of executed methods after execution filtering.}
    \label{fig:method_filtering_compression}
    
\end{figure}
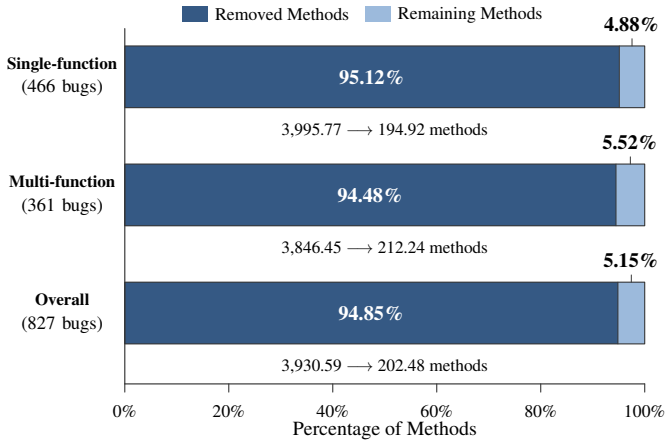

\textbf{Execution Filtering.} As shown in Fig.~\ref{fig:method_filtering_compression}, EF reduces the average number of executed methods from 3,930.59 to 202.48, removing 94.85\% of the original methods. Single-function and multi-function bugs show similar reductions of 95.12\% and 94.48\%, respectively. These results indicate that failing tests exercise only a small fraction of repository methods, and EF substantially narrows subsequent dynamic analysis by removing unexecuted paths.

\begin{figure}[t]
    \centering
    \includegraphics[width=\columnwidth]{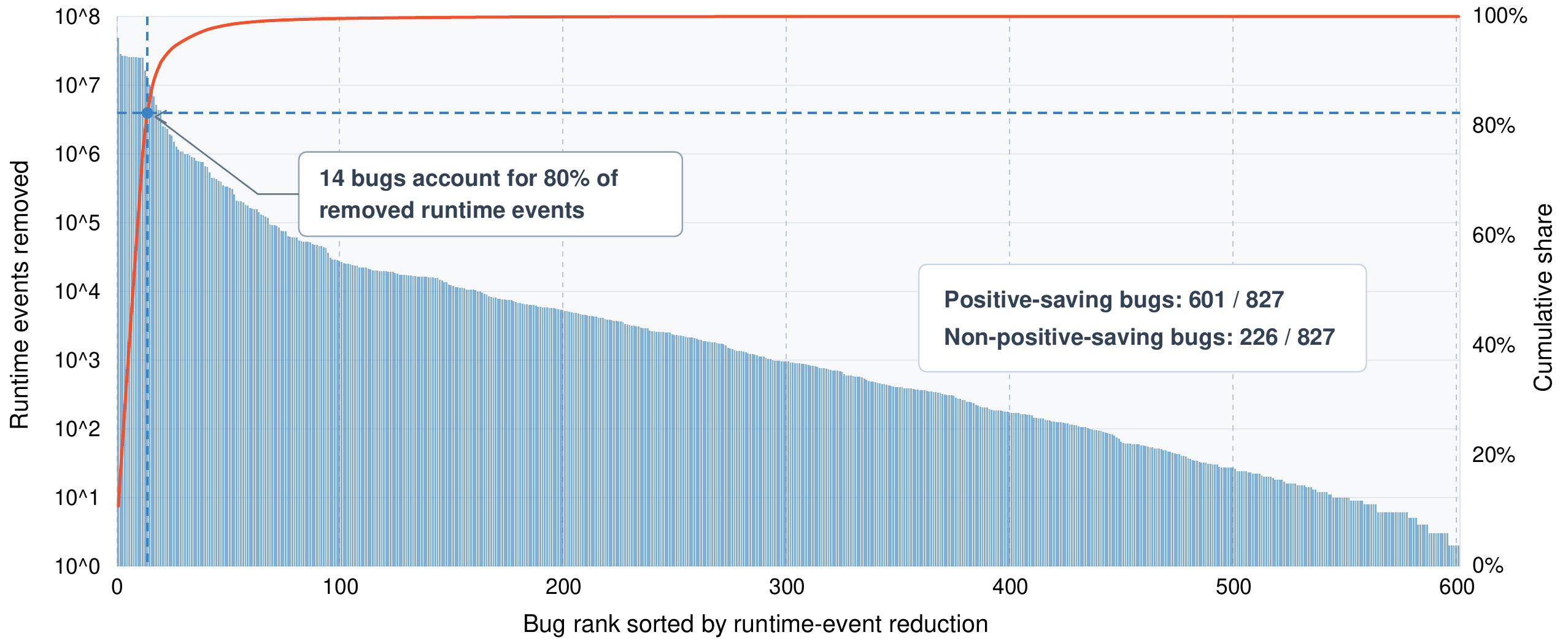}
    \caption{Ranked distribution of runtime-event reductions introduced by structural filtering.}
    \label{fig:rq2_trace_event_pareto}
    
\end{figure}

\textbf{Structural Filtering.} Building on EF, we evaluate how SF affects dynamic-information size. As shown in Table~\ref{tab:structural_filtering_reduction}, SF reduces the average number of executed methods from 202.48 to 168.22 (16.9\% reduction), runtime events from 1,305.28K to 816.26K (37.5\% reduction), and trace size from 506.33 MB to 332.57 MB (34.3\% reduction). Runtime events and trace size decrease more sharply than the number of methods because the low-value methods removed by SF, such as getters and simple delegates, are limited in number but frequently invoked. As shown in Fig.~\ref{fig:rq2_trace_event_pareto}, SF reduces runtime events for 601 of 827 bugs, while only 14 bugs account for 80\% of the removed events. This indicates that frequent invocations of low-value methods can cause extreme trace growth for some bugs. Without SF, such redundant events would obscure behaviors related to fault propagation and reduce the usability of dynamic information. SF therefore reduces the average trace size while preventing extreme traces from degrading dynamic analysis.

\textbf{Behavior Filtering.} After EF and SF, BF further removes redundant records at the behavioral level. As shown in Table~\ref{tab:behavior_filtering}, BF reduces retained records from 473.87K to 208.64K (55.97\% reduction) and trace size from 247.18 MB to 133.75 MB (45.89\% reduction). Across bugs, BF reduces retained records for 824 of 827 bugs; 660 bugs show reductions of at least 25\%, including 274 with reductions of at least 50\% and 75 with reductions above 75\%. These results show that BF consistently removes redundant dynamic records across most bugs rather than benefiting only a few extreme traces.

\begin{table}[t]
    \centering
    \caption{Reduction achieved by structural filtering.}
    \label{tab:structural_filtering_reduction}

    \renewcommand{\arraystretch}{1.18}
    \setlength{\tabcolsep}{9pt}

    \resizebox{\columnwidth}{!}{%
        \begin{tabular}{lrrr}
            \toprule
            \textbf{Metric}
            & \textbf{EF only}
            & \textbf{EF + SF}
            & \textbf{Reduction} \\
            \midrule

            Executed methods per bug
            & 202.48
            & 168.22
            & $-16.9\%$ \\

            Runtime events per bug (K)
            & 1305.28
            & 816.26
            & $-37.5\%$ \\

            Trace volume per bug (MB)
            & 506.33
            & 332.57
            & $-34.3\%$ \\

            \bottomrule
        \end{tabular}%
    }

\end{table}

\begin{table}[t]
    \centering
    \caption{Effectiveness of behavior filtering.}
    \label{tab:behavior_filtering}

    \renewcommand{\arraystretch}{1.2}
    \setlength{\tabcolsep}{5pt}

    \begin{minipage}{\columnwidth}
    \centering
    \resizebox{\columnwidth}{!}{%
    \begin{tabular}{l c c c r c c r}
        \toprule

        \multirow{2}{*}{\textbf{Bug Type}} &
        \multirow{2}{*}{\textbf{\#Bugs}} &
        \multicolumn{3}{c}{\textbf{Records (K)}} &
        \multicolumn{3}{c}{\textbf{Trace Size (MB)}} \\

        \cmidrule(lr){3-5}
        \cmidrule(lr){6-8}

        &
        &
        {\footnotesize\textbf{EF + SF}} &
        {\footnotesize\textbf{EF + SF + BF}} &
        {\footnotesize\textbf{Reduction}} &
        {\footnotesize\textbf{EF + SF}} &
        {\footnotesize\textbf{EF + SF + BF}} &
        {\footnotesize\textbf{Reduction}} \\

        \midrule

        Single-function &
        466 &
        495.52 &
        211.94 &
        $-57.23\%$ &
        276.89 &
        149.26 &
        $-46.09\%$ \\

        Multi-function &
        361 &
        445.92 &
        204.37 &
        $-54.17\%$ &
        208.82 &
        113.72 &
        $-45.54\%$ \\

        ALL &
        827 &
        473.87 &
        208.64 &
        $-55.97\%$ &
        247.18 &
        133.75 &
        $-45.89\%$ \\

        \bottomrule
    \end{tabular}
    }

    \end{minipage}
    
\end{table}

\noindent\textbf{Answer to RQ2.}
TEG progressively removes irrelevant and redundant dynamic information through EF, SF, and BF, producing compact and queryable runtime evidence for repair reasoning.

\subsection{RQ3: Reasoning Diversity and Repair Complementarity}
\label{sec:rq3}

To answer RQ3, we analyze whether multi-perspective agents can introduce diversity and complementarity into repair on all 854 bugs. At the reasoning level, we compare the reasoning diversity of CT-Repair and ReinFix. To avoid bias from method names and presentation order in LLM-based evaluation, we anonymize the two methods, randomly shuffle their order, and ask three LLM judges to independently evaluate each comparison three times, reporting the mean and standard deviation. At the patch level, we use the exact duplication rate to measure the proportion of identical patches and normalized AST distance to measure structural differences between patches, and compare patch diversity within the same agent (Intra-Agent) and across different agents (Inter-Agent). At the repair level, we report the bugs correctly repaired by Agent-S, Agent-D, and Agent-H, as well as their union and overlap, to assess whether multi-perspective reasoning expands repair coverage.

\begin{table}[t]
    \centering
    \caption{
        Reasoning-level evidence diversity comparison between CT-Repair and ReinFix.
    }
    \label{tab:reasoning_diversity}

    \renewcommand{\arraystretch}{1.20}
    \setlength{\tabcolsep}{11pt}

    \resizebox{\columnwidth}{!}{
    \begin{tabular}{@{}l l c c c@{}}
        \toprule
        \textbf{Metric}
        & \textbf{LLM Judge}
        & \textbf{CT-Repair}
        & \textbf{ReinFix}
        & \textbf{Difference} \\
        \midrule

        \multirow{3}{*}{Evidence cluster count $\uparrow$}
        & GPT-5.5
        & $\mathbf{1.695} \pm \mathbf{0.008}$
        & $1.253 \pm 0.012$
        & $+0.442$ \\

        & Gemini-3.5-flash
        & $\mathbf{1.674} \pm \mathbf{0.004}$
        & $1.143 \pm 0.005$
        & $+0.531$ \\

        & DeepSeek-V4-Pro
        & $\mathbf{1.683} \pm \mathbf{0.015}$
        & $1.048 \pm 0.004$
        & $+0.635$ \\

        \midrule

        \multirow{3}{*}{Perspective diversity score $\uparrow$}
        & GPT-5.5
        & $\mathbf{2.172} \pm \mathbf{0.013}$
        & $1.401 \pm 0.014$
        & $+0.771$ \\

        & Gemini-3.5-flash
        & $\mathbf{2.235} \pm \mathbf{0.014}$
        & $1.240 \pm 0.004$
        & $+0.995$ \\

        & DeepSeek-V4-Pro
        & $\mathbf{2.327} \pm \mathbf{0.025}$
        & $1.094 \pm 0.009$
        & $+1.233$ \\

        \midrule

        \multirow{3}{*}{Duplicate evidence count $\downarrow$}
        & GPT-5.5
        & $\mathbf{1.259} \pm \mathbf{0.008}$
        & $1.641 \pm 0.012$
        & $-0.382$ \\

        & Gemini-3.5-flash
        & $\mathbf{1.281} \pm \mathbf{0.004}$
        & $1.751 \pm 0.005$
        & $-0.470$ \\

        & DeepSeek-V4-Pro
        & $\mathbf{1.271} \pm \mathbf{0.015}$
        & $1.845 \pm 0.004$
        & $-0.574$ \\

        \bottomrule
    \end{tabular}
    }
    \vspace{-5pt}
\end{table}

\textbf{Reasoning diversity of multi-perspective Agents.} As shown in Table~\ref{tab:reasoning_diversity}, CT-Repair produces more diverse evidence than ReinFix under all three LLM judges. Compared with ReinFix, CT-Repair increases the number of evidence clusters by 0.44--0.64 and the perspective diversity score by 0.77--1.23, while reducing repeated evidence by 0.38--0.57. The consistent evaluations show that Agent-S, Agent-D, and Agent-H derive distinct root-cause analyses and repair strategies from different information sources rather than repeatedly sampling the same repair process.

\textbf{Diversity of the candidate patch space.} As shown in Table~\ref{tab:patch_level_diversity}, for single-function bugs, the exact duplication rate decreases from 30.68\% to 22.83\%, while the median normalized AST distance increases from 0.0397 to 0.0681, a 71.5\% improvement. For multi-function bugs, the duplication rate decreases from 28.61\% to 19.43\%, while the median normalized AST distance increases from 0.0405 to 0.0794, a 96.0\% improvement. These results show that multi-perspective reasoning transforms distinct fault understandings into a broader candidate patch space, with a stronger effect on multi-function bugs involving cross-method dependencies.

\begin{table}[t]
    \centering
    \caption{Patch-level diversity of intra-agent and inter-agent candidates.}
    \label{tab:patch_level_diversity}

    \renewcommand{\arraystretch}{1.15}
    \setlength{\tabcolsep}{10pt}

    \resizebox{\columnwidth}{!}{
    \begin{tabular}{@{}llccc@{}}
        \toprule
        \textbf{Setting}
        & \textbf{Comparison}
        & \textbf{Dup. Rate $\downarrow$}
        & \textbf{AST Distance $\uparrow$}
        & \textbf{AST Gain $\uparrow$} \\
        \midrule

        \multirow{2}{*}{Single-function}
        & Intra-Agent
        & 30.68\%
        & 0.0397
        & -- \\

        & Inter-Agent
        & \textbf{22.83\%}
        & \textbf{0.0681}
        & \textbf{+71.5\%} \\

        \midrule

        \multirow{2}{*}{Multi-function}
        & Intra-Agent
        & 28.61\%
        & 0.0405
        & -- \\

        & Inter-Agent
        & \textbf{19.43\%}
        & \textbf{0.0794}
        & \textbf{+96.0\%} \\

        \bottomrule
    \end{tabular}
    }

    \vspace{2pt}

    \begin{minipage}{\columnwidth}
    \tiny
    \raggedright
    \textit{Note:}
    Dup. Rate is the exact duplicate rate of patch pairs. AST Distance is the median normalized AST distance between patches. AST Gain measures the relative improvement of Inter-Agent over Intra-Agent. Arrows indicate better directions.
    \end{minipage}

\end{table}

\begin{table}[t]
    \centering
    \caption{Repair overlap and complementarity among evidence-perspective agents.}
    \label{tab:agent_complementarity}

    \footnotesize
    \renewcommand{\arraystretch}{1.08}
    \setlength{\tabcolsep}{0pt}

    \begin{tabular*}{\columnwidth}{@{\extracolsep{\fill}}lcccc@{}}
        \toprule
        \textbf{Subset}
        & \textbf{Agent-S}
        & \textbf{Agent-D}
        & \textbf{Agent-H}
        & \textbf{\#Fixes} \\
        \midrule
        S only
        & \(\bullet\)
        & \(\circ\)
        & \(\circ\)
        & 53 \\

        D only
        & \(\circ\)
        & \(\bullet\)
        & \(\circ\)
        & 29 \\

        H only
        & \(\circ\)
        & \(\circ\)
        & \(\bullet\)
        & 35 \\

        S \(\cap\) D only
        & \(\bullet\)
        & \(\bullet\)
        & \(\circ\)
        & 30 \\

        S \(\cap\) H only
        & \(\bullet\)
        & \(\circ\)
        & \(\bullet\)
        & 51 \\

        D \(\cap\) H only
        & \(\circ\)
        & \(\bullet\)
        & \(\bullet\)
        & 35 \\

        S \(\cap\) D \(\cap\) H
        & \(\bullet\)
        & \(\bullet\)
        & \(\bullet\)
        & 256 \\
        \midrule
        \textbf{Total}
        & \textbf{390}
        & \textbf{350}
        & \textbf{377}
        & \textbf{489} \\
        \bottomrule
    \end{tabular*}

    \vspace{2pt}

    \begin{minipage}{\columnwidth}
        \tiny
        \raggedright
        \textit{Note:}
        \(\bullet\) indicates that the agent repairs the subset, and \(\circ\) otherwise.
    \end{minipage}
\end{table}

\textbf{Repair complementarity across agent perspectives.} As shown in Table~\ref{tab:agent_complementarity}, Agent-S, Agent-D, and Agent-H correctly repair 390, 350, and 377 bugs, respectively, while their union reaches 489. Compared with the best individual agent, Agent-S, combining the three perspectives repairs 99 additional bugs, a 25.38\% improvement. The agents uniquely repair 53, 29, and 35 bugs, respectively, and jointly repair 256 bugs. These results show that the three perspectives are not redundant; each covers distinct bug scenarios, and their combination transforms diverse fault understandings into complementary repair strategies, substantially expanding repair coverage.

\noindent\textbf{Answer to RQ3.}
Multi-perspective agents improve reasoning diversity, reduce patch duplication, increase structural differences, and expand repair coverage beyond repeated sampling within a single perspective.

\subsection{RQ4: Contribution of Main Components}
\label{sec:rq4}

To answer RQ4, we conduct a component-wise ablation study to evaluate each major component of CT-Repair. Due to the high computational cost of performing a full ablation on the complete benchmark, we primarily conduct this study on the 395 bugs from Defects4J v1.2. Using the full configuration as the baseline, we separately remove CPG, TEG, Agent-S, Agent-D, Agent-H, and iterative refinement, and compare changes in the number of correctly repaired bugs.

\begin{table}[t]
    \centering
    \caption{Ablation study of CT-Repair components on Defects4J v1.2.}
    \label{tab:ablation_results}

    \renewcommand{\arraystretch}{1.18}
    \setlength{\tabcolsep}{16pt}

    \resizebox{\columnwidth}{!}{%
    \begin{tabular}{lcccc}
        \toprule
        \textbf{Method}
        & \textbf{Multi-function}
        & \textbf{Single-function}
        & \textbf{All}
        & \textbf{$\Delta$ Impact (\%)} \\
        \midrule

        w/o TEG
        & 53
        & 171
        & 224
        & $-8.6\%$ \\

        w/o CPG
        & 52
        & 164
        & 216
        & $-11.8\%$ \\

        w/o Agent-S
        & 52
        & 163
        & 215
        & \textbf{$-12.2\%$} \\

        w/o Agent-D
        & 54
        & 177
        & 231
        & $-5.7\%$ \\
        
        w/o Agent-H
        & 54
        & 169
        & 223
        & $-9.0\%$ \\

        w/o Iteration
        & 57
        & 176
        & 233
        & $-4.9\%$ \\

        \midrule
        \textbf{CT-Repair$_{\mathit{Full}}$}
        & \textbf{65}
        & \textbf{180}
        & \textbf{245}
        & -- \\
        \bottomrule
    \end{tabular}%
    }
\end{table}

\textbf{Ablation Results.} As shown in Table~\ref{tab:ablation_results}, removing any component degrades CT-Repair's repair performance. Removing CPG and TEG reduces correctly repaired bugs from 245 to 216 and 224, corresponding to drops of 11.8\% and 8.6\%, respectively. This indicates that static structural information and dynamic execution information jointly support fault understanding, with CPG capturing code structure and dependencies and TEG providing runtime state changes and fault propagation information. Removing Agent-S, Agent-H, and Agent-D reduces correctly repaired bugs to 215, 223, and 231, corresponding to drops of 12.2\%, 9.0\%, and 5.7\%, respectively. Although the three perspectives contribute differently, each independently improves the overall repair capability. Finally, disabling iterative refinement reduces correctly repaired bugs to 233, a drop of 4.9\%, showing that validation-feedback-based strategy adjustment further improves patch quality.

\noindent\textbf{Answer to RQ4.}
The ablation study shows that CT-Repair's effectiveness comes from the combined contributions of CPG, TEG, multi-perspective agents, and iterative refinement.

\section{Threats to Validity}

\textbf{Data Leakage.} 
LLM training data may contain correct patches for benchmark bugs, affecting evaluation objectivity. To mitigate this risk, we reproduced ReinFix and RepairAgent using the same base model in Section~\ref{sec:rq1}, where CT-Repair still achieves better repair effectiveness. The ablation study in Section~\ref{sec:rq4} further verifies the consistent contributions of all main components. Nevertheless, we cannot completely rule out data leakage. In the future, we plan to further evaluate CT-Repair on real-world bugs collected after the models' training-data cutoff dates. 

\textbf{Evaluation Assumptions.}
Our evaluation relies on two assumptions that may affect external validity. First, following recent APR studies~\cite{xia2024automated,yin2024thinkrepair,zhang2025repair}, CT-Repair is evaluated under perfect fault localization. This setting aligns our evaluation with existing baselines and avoids additional bias from different FL tools, but may overestimate CT-Repair's performance in realistic end-to-end repair scenarios. Second, Section~\ref{sec:rq3} relies on LLM judges to assess reasoning diversity, potentially introducing bias. To mitigate this risk, we anonymize method names, use multiple LLM judges, and repeat the experiments. We also manually inspected 100 sampled cases and found strong agreement between human labels and aggregated LLM judgments (Cohen's $\kappa = 0.82$). Future work will combine CT-Repair with realistic FL techniques and conduct larger-scale human annotation.

\textbf{Repair Costs.}
LLM-based repair inevitably incurs additional inference costs. Due to the mixed-model configuration, CT-Repair's monetary cost is difficult to compare directly, so we focus on average token consumption. As shown in Table~\ref{tab:llm_settings} of Section~\ref{sec:rq1}, CT-Repair consumes 178.35K tokens per bug on average, lower than ChatRepair (467K), AdverIntent-Agent (438K), and RepairAgent (270K). Notably, both CT-Repair and RepairAgent use an FSM to guide repair, while CT-Repair reduces token consumption by 33.9\%. This suggests that the summary-based state management mechanism introduced in Section~\ref{sec:fsm_reasoning} effectively reduces cross-state context redundancy. CT-Repair also reduces the average runtime from 920.0 s for RepairAgent to 158.9 s. Given the different base models and runtime environments, this runtime comparison merely reflects differences in overall cost magnitude.

\section{Related Work}

APR research has entered the era of LLMs~\cite{zhang2024systematic}. AlphaRepair~\cite{xia2022less} and GAMMA~\cite{zhang2023gamma} use LLMs through cloze-style generation or template-based prompting. ChatRepair~\cite{xia2024automated} and Self-Debugging~\cite{chen2024teaching} further feed test failures or execution feedback back to the model to iteratively refine patches.

Researchers have further introduced dynamic execution information into LLM-based APR. TraceFixer~\cite{bouzenia2023tracefixer} adds local variable values and expected execution states during CodeT5 fine-tuning; LDB~\cite{zhong2024debug} tracks basic-block-level intermediate variable values and uses LLMs to validate program states step by step; DynaFix~\cite{huang2025dynafix} instruments programs to collect variable states, control-flow paths, and call stacks for iterative repair. These studies show that static and dynamic evidence can support LLM-based repair. However, existing methods mostly provide such evidence as textual context, whereas CT-Repair organizes it into queryable CPGs and TEGs to support hypothesis-driven evidence retrieval.

Recent studies have shifted from one-shot patch generation to agentic repair based on feedback and tool interaction. Existing agent-based methods can be broadly grouped into single-agent tool-augmented repair and multi-agent collaborative repair. Among single-agent methods, ReinFix~\cite{zhang2025repair}, RepairAgent~\cite{bouzenia2025repairagent}, SWE-agent~\cite{yang2024swe}, AutoCodeRover~\cite{zhang2024autocoderover}, and SpecRover~\cite{ruan2025specrover} support fault localization, root-cause analysis, and patch generation through repair-ingredient retrieval, repository navigation, code search, context retrieval, or reproduction tests. Among multi-agent methods, UniDebugger~\cite{lee2025unidebugger}, MAGIS~\cite{tao2024magis}, SWE-Debate~\cite{li2025swe}, and TraceRepair~\cite{wu2026runtime} guide the repair workflow and validate or refine patches through role specialization, debate mechanisms, or runtime constraints. These methods mainly emphasize tool use, workflow collaboration, and patch selection. In contrast, CT-Repair focuses on deriving diverse repair strategies from distinct evidence perspectives, allowing agents to analyze faults independently from static, dynamic, and hybrid evidence before patch generation and thereby expanding the repair-strategy space.

\section{Conclusion}
We presented CT-Repair, an agentic APR framework that combines queryable static and dynamic program representations with multi-perspective reasoning. CT-Repair uses a CPG to represent structural dependencies and a TEG to organize runtime behavior after multi-stage filtering. Static-oriented, dynamic-oriented, and hybrid-oriented agents independently derive root-cause hypotheses and repair strategies, thereby improving the diversity of both reasoning processes and candidate patches. Extensive experiments show that CT-Repair effectively compresses redundant dynamic information, broadens the repair-strategy space, and improves the program repair capability of LLMs while maintaining reasonable repair costs.

\section{Data Availability}
\label{sec:data}

To support independent verification and replication, we provide anonymized research artifacts for CT-Repair at \url{https://anonymous.4open.science/r/CT-Repair-6D29/}.

\bibliographystyle{IEEEtran}
\bibliography{ref}

\end{document}